\documentclass[aps,pra,superscriptaddress,amsmath,amssymb,preprintnumbers]{revtex4}
\usepackage{amssymb}
\usepackage{booktabs}
\usepackage{graphicx}
\usepackage{bm}
\usepackage{color}
\usepackage{cancel}
\usepackage{slashed}
\usepackage{subfigure}

\newcommand{\Ignore}[1]{}

\newcommand{\Ket}[1]{\left\vert #1\right\rangle}
\newcommand{\Bra}[1]{\left\langle #1\right\vert}

\newcommand{\bathsize}{{L}}

\begin{document}

\pagecolor{white}

\title{Adiabatic manipulation of a system interacting with a spin-bath}

\author{Benedetto Militello}
\author{Anna Napoli}
\affiliation{Universit\`a degli Studi di Palermo, Dipartimento di Fisica e Chimica - Emilio Segr\`e, Via Archirafi 36, 90123 Palermo, Italia}
\affiliation{I.N.F.N. Sezione di Catania, Via Santa Sofia 64, I-95123 Catania, Italia}



%
%
%
%
%

\begin{abstract}
Stimulated Raman Adiabatic Passage, a very efficient technique for manipulating a quantum system based on the adiabatic theorem, is analyzed in the case where the manipulated physical system is interacting with a spin bath. Exploitation of the rotating wave approximation allows for the identification of a constant of motion which simplifies both the analytical and the numerical treatment, which allows for evaluating the total unitary evolution of system and bath. The efficiency of the population transfer process is investigated in several regimes, including the weak and strong coupling with the environment and the off-resonance. The formation of appropriate Zeno subspaces explains the lowering of the efficiency in the strong damping regime.
\end{abstract}

\maketitle


\section{Introduction}\label{sec:introduction}

The adiabatic theorem~\cite{ref:Messiah,ref:Griffiths}, according to which a quantum system subjected to a slow varying Hamiltonian evolves in such a way to keep constant the populations of the instantaneous eigenstates of the Hamiltonian (the so called adiabatic following of the eigenstates of the Hamiltonian), is a key tool in quantum technologies. Indeed, it allows for efficiently manipulating the state of the relevant quantum system~\cite{ref:Lidar2018,ref:Santos2020,ref:Ilin2021,ref:Pys2022,ref:Coello2022}.
Stimulated Raman Adiabatic Passage (STIRAP)~\cite{ref:STIRAP_rev1,ref:STIRAP_rev2,ref:STIRAP_rev3,ref:STIRAP_rev4,ref:STIRAP_rev5} is a widely used technique lying in the realm of adiabatic evolutions. The standard scheme is aimed at realizing a complete population transfer from one state to another, which is realized by exploitation of a Raman coupling scheme involving an auxiliary state coupled to the other two through suitable pulses having time-dependent amplitudes. The specific features of such pulses should ensure the validity of the adiabatic approximation and make an eigenstate of the time-dependent Hamiltonian coincide with the initial state of the system at the beginning of the process and with the target state at the end of the application of the pulses. This technique is still extensively studied~\cite{,ref:Cinins2022,ref:Liu2023,ref:Genov2023} and improvements of the original scheme have been proposed, for example including shortcuts to adiabaticity, which allow for shortening the duration of the process~\cite{ref:Guery2019,ref:Stefanatos2020,ref:Stefanatos2022,ref:Messikh2022,ref:Evange2023}, though  this modified technique has the disadvantage of requiring a more complicated apparatus to implement additional Hamiltonian terms necessary for the shortcuts.

Since no system is perfectly isolated neither an experimental setup can be lacking imperfections, the stability of the process with respect to uncertainty or fluctuations of the relevant parameters, such as amplitudes and phases of the pulses, has been studied~\cite{ref:Genov2013,ref:Yatsenko2014}. Also the role of quantum noise associated with the presence of the quantized electromagnetic fields has been extensively investigated with different approaches, including non-Hermitian Hamiltonians, which is appropriate in the case where some of the involved state is undergoing decays toward states different from those involved in the STIRAP scheme~\cite{ref:Vitanov1997}. When the decays happen to induce transitions between the states of the STIRAP scheme, the non-Hermitian Hamiltoninan approach does not fit anymore since it cannot describe incoherent transitions between the states of the Raman scheme. Therefore, a master equation approach is necessary, and the relevant equation in the Lindblad form can be obtained under different assumptions~\cite{ref:Ivaniv2005,ref:Scala2010,ref:Scala2011,ref:PhysScr2011,ref:Mathisen2018}, according to the general theory of open quantum systems in the Markovian limit~\cite{ref:Petruccione,ref:Gardiner}, starting with a Hamiltonian interaction model between the three-state system and a bosonic bath corresponding to the electromagnetic field. 

The physical scenarios of interest for the STIRAP technique range from cold gases~\cite{ref:Ni2008,ref:Danzl2008,ref:Danzl2010} to condensed matter~\cite{ref:Halfmann2007,ref:Alexander2008,ref:Golter2014,ref:Yale2016,ref:Baksic2016,ref:Zhou2017,ref:Wolfowicz2021}, plasmonic systems~\cite{ref:Varguet2016,ref:Castellini2018}, superconducting devices~\cite{ref:Kubo2016,ref:Xu2016,ref:Kumar2016} and trapped ions~\cite{ref:Soresen2006,ref:Higgins2017}. Though the typical source of quantum noise considered in such systems is due to radiative processes associated to the interaction with the electromagnetic field, in many cases the system to be manipulated is close enough to other atomic systems to be considered in interaction with them. This is the case for example of Nitrogen Vacancies in diamond~\cite{{ref:Golter2014,ref:Baksic2016,ref:Zhou2017}} or rare-earth doped crystals~\cite{ref:Halfmann2007,ref:Alexander2008}.
Recently, adiabatic manipulation of spin defects in magnetics materials has been studied, including the environmental effects due to the interaction with the surrounding spins~\cite{ref:Onizhuk2021}. It is then meaningful to consider STIRAP processes in the presence of interaction with a spin environment, which fits, for example, the physical scenarios of solid state physics and nuclear magnetic resonance (NMR).

In this paper, we present a theoretical and numerical study of the effects of a spin bath made of identical particles on a quantum system manipulated through STIRAP technique. Though descriptions of system interacting with spin environments through master equations are possible~\cite{ref:Fischer2007,ref:Ferraro2008,ref:Bhatta2017}, we will consider unitary dynamics.  Indeed, thanks to the rotating wave approximation (RWA) and the consequent conservation of the number of excitations, the total effective dynamics of system and bath can be analyzed, making some specific analytical predictions and efficient numerical simulations. We first analyze theoretically the model, predicting some general behaviors. Second, focusing on the homogeneous case which constitutes a significant mathematical simplification, we  report on some numerical simulations confirming the analytical predictions. Such simplified version of the model emerges when all the spins interact with the three-state system in the same way. Though these assumptions appear somehow restrictive, there are experimental scenarios where it has been realized, such as NMR spectroscopy~\cite{ref:Joshi2022}.

The paper is structured as follows. In sec.~\ref{sec:PhysicalSystem} the Hamiltonian model describing the three-state system interacting with a spin environment is introduced. In sec.~\ref {sec:theoretical} an analytical study of the model is presented in order to predict the behavior of the efficiency in some special regimes, such as the weak and the strong damping. Sec.~\ref {sec:numerics} is dedicated to the numerical results obtained for the homogeneous model. Finally, in sec.~\ref{sec:discussion} we extensively discuss the numerical and analytical results.


\section{Physical Model}\label{sec:PhysicalSystem}

\begin{figure}[h]
\begin{center}
\includegraphics[width=0.45\textwidth, angle=0]{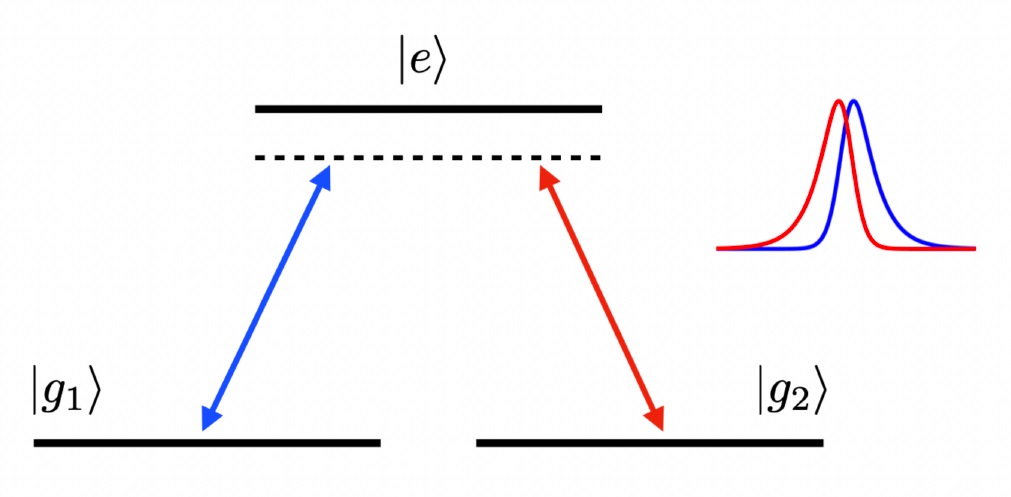}
\end{center}
\caption{$\Lambda$-coupling scheme involving the three atomic states $\Ket{g_1}$, $\Ket{g_2}$ and $\Ket{e}$. Population is supposed to be transferred from state $\Ket{g_1}$ to state $\Ket{g_2}$, which is accomplished via two pulses in the so called counterintuitive sequence, where the coupling between $\Ket{g_2}$ and $\Ket{e}$ precedes the coupling between $\Ket{e}$ and $\Ket{g_1}$. The inset represents the typical shape of the pulses.} \label{fig:Lambda_scheme}
\end{figure}

{ \em STIRAP Hamiltonian---} Let us consider a three-state system: one state, $\Ket{g_1}$, is assumed as the initial condition, another one, $\Ket{g_2}$, is the target state, where population should be transferred from the initial state, and the third one, $\Ket{e}$, is an auxiliary state. As shown in Fig.~\ref{fig:Lambda_scheme}, each of the first two states is coupled to the third one by two coherent pulses, which allows for the realization of an adiabatic following responsible for the required population transfer. Here we assume that $\Ket{g_1}$ and $\Ket{g_2}$ belong to a degenerate energy subspace, while the auxiliary state $\Ket{e}$ has an energy gap $\nu$ with the other two.

This three-state system in the presence of the pulses is governed by the Hamiltonian ($\hbar=1$): 
\begin{eqnarray}\label{eq:HS}
\nonumber
H_S &=& \nu \Ket{e}\Bra{e} 
+ \Omega_p(t) (e^{i\nu't} \Ket{g_1}\Bra{e} + e^{-i\nu't} \Ket{e}\Bra{g_1} ) \\
&+& \Omega_s(t) (e^{i\nu't} \Ket{g_2}\Bra{e} + e^{-i\nu't} \Ket{e}\Bra{g_2})\,.
\end{eqnarray}
In the rotating frame whose relevant transformation is generated by $G_S=\nu' \Ket{e}\Bra{e}$, it becomes:
\begin{eqnarray}\label{eq:HS-tilde}
\tilde{H}_S &=& \Delta_S \Ket{e}\Bra{e}  + \Omega_p(t) (\Ket{g_1}\Bra{e} + \Ket{e}\Bra{g_1} )
+ \Omega_s(t) (\Ket{g_2}\Bra{e} + \Ket{e}\Bra{g_2})\,,
\end{eqnarray}
where $\Delta_S = \nu-\nu'$. 
The instantaneous eigenvalues of this operator are $0, \, (\Delta_S \pm (\Delta_S^2 + 4 (\Omega_s^2 + \Omega_p^2))^{1/2} ) / 2$, and the eigenstate corresponding to zero is:
\begin{equation}\label{eq:DarkState}
\Ket{0}= \cos\theta\Ket{g_1}-\sin\theta\Ket{g_2}\,, \quad \tan\theta(t)=\frac{\Omega_p(t)}{\Omega_s(t)} \\
\end{equation}

The standard and generally more efficient way to transfer population from $\Ket{g_1}$ to $\Ket{g_2}$ is exploitation of the so called counterintuitive sequence, consisting in applying the two pulses in such a way that $\Omega_s$ precedes $\Omega_p$. It implies that $\theta$ initially assumes the value $0$ and eventually the value $\pi/2$, which in turn implies that the state $\Ket{0}$ initially coincides with $\Ket{g_1}$ and finally with $\Ket{g_2}$.
A typical choice for the pulses is the following:
\begin{subequations}
\begin{eqnarray}
\label{eq:pulse_s}
\Omega_s &=&  \frac{\Omega_0}{\sqrt{2}} \mathrm{sech}\left( t/\tau \right) \cos\left[ \pi/4 (\mathrm{tanh}(t/\tau)+1)\right] \,, \\
\label{eq:pulse_p}
\Omega_p &=&  \frac{\Omega_0}{\sqrt{2}} \mathrm{sech}\left( t/\tau \right) \sin\left[ \pi/4 (\mathrm{tanh}(t/\tau)+1)\right] \,.
\end{eqnarray}
\end{subequations}
Though the analytic expression of such functions could seem weird, the relevant edges reproduce pretty well the typical shapes of pulses in real experiments and, on the other hand, allow for obtaining analytical results in many cases~\cite{ref:Vitanov1997}.
Time evolution occurs in a time-window $[-T, T]$ large enough to have the amplitudes of the two pulses essentially vanish at $t=\pm T$. The width ($\tau$) and amplitude ($\Omega_0$) of the pulses must be chosen in such a way to satisfy the condition for the adiabatic approximation. This means that the ratios between matrix elements of the time derivative of the Hamiltonian and the squares of the relevant transition frequencies are supposed to be much smaller than unity. Specifically, since such ratios are proportional to $\Omega_0 \tau$, it is sufficient to require $(\Omega_0 \tau)^{-1} \ll 1$.

{ \em System-environment interaction ---} Since the three-state system is interacting with its environment, the relevant free and interaction Hamiltonian must be considered, starting from the Schr\"odinger picture and then moving again to the rotating frame previously mentioned. Assuming that the environment consists of a large set of identical two-state systems, the total Hamiltonian is the following:
\begin{equation}\label{eq:H0}
H = H_S + H_E + H_I\,,
\end{equation}
where $H_S$ is the free Hamiltonian of the system given in \eqref{eq:HS}, while
\begin{equation}\label{eq:HE}
H_E = \sum_{k=1}^\bathsize \frac{\omega}{2} (\sigma_z^{(k)} + 1_k) \,,
\end{equation}
is the free Hamiltonian of the environment, with $L$ the number of spins (or, more generally, two-state systems) and $\omega$ the natural frequency of such spins (whose energy in their state $\Ket{\downarrow}$ is assumed to be zero). Finally, the system-environment coupling is given by:
\begin{eqnarray}\label{eq:HI}
H_I &=& \sum_{m=1,2} \left(\Ket{g_m}\Bra{e}+\Ket{e}\Bra{g_m}\right)\otimes\sum_{k=1}^\bathsize \frac{\eta_k^{(m)}}{2} \sigma_x^{(k)} \,, \,\,\,\,\,\,\,
\end{eqnarray}
where the real numbers $\eta_k^{(m)}$ is the coupling constant between the $k$-th spin and the $\Ket{g_m}-\Ket{e}$ transition. Here $\sigma_k^{(\alpha)}$ are the Pauli operators associated to the $k$-th spin while $1_k$ is the relevant identity operator. 

In the rotating frame, whose relevant transformation is generated by $G_S + G_E$, with $G_E = (\nu'/2) \sum_k \sigma_z^{(k)}$, we get: 
\begin{equation}\label{eq:HE-tilde}
\tilde{H}_E = \sum_{k=1}^\bathsize \frac{\Delta_E}{2} (\sigma_z^{(k)} + 1_k) \,,
\end{equation}
with $\Delta_E = \omega - \nu'$, and the following system-bath interaction term:
\begin{eqnarray}\label{eq:HI-tilde}
\tilde{H}_I &=& \frac{1}{2} \sum_{m=1,2} \Ket{g_m}\Bra{e} \otimes\sum_{k=1}^\bathsize \eta_k^{(m)} \sigma_+^{(k)} + H.c.\,, 
\end{eqnarray}
which has been evaluated under the RWA, which means neglecting the so called counterrotating terms, that in our case are $2^{-1} \sum_{m,k}\Ket{g_m}\Bra{e} \eta_k^{(m)} \sigma_-^{(k)}+H.c.$. This approximation is crucial in our treatment since it allows the identification of a constant quantity, as it is well clarified in the following.

All considered,  the total Hamiltonian in the rotating frame is $\tilde{H}_S + \tilde{H}_E + \tilde{H}_I$.

{ \em Homogeneous model---} The homogeneous model emerges when all the spins of the bath are assumed to interact equally with the three-state system ($\forall k$ $\eta_k^{(1)} = \eta_k^{(2)} \equiv \eta$). In this case, after introducing $J_\alpha = \sum_{k=1}^\bathsize \sigma_\alpha^{(k)}/2$, we get:
\begin{eqnarray}
\tilde{H}_{E} &=& \Delta_E (J_z + L/2)\,,
\label{eq:HE-tilde}
\end{eqnarray}
\begin{eqnarray}\label{eq:HI-tilde-rwa}
\tilde{H}_{I} &=& \eta \left(\Ket{e}\Bra{g_1} +\Ket{e}\Bra{g_2} \right) \otimes J_- + H.c. \,.
\end{eqnarray}
From these expressions, it naturally emerges that the most appropriate basis to describe the environment in this case is the basis of the total angular momentum $\Ket{j,m}$.

\begin{figure}[h!]
\begin{center}
\includegraphics[width=0.45\textwidth, angle=0]{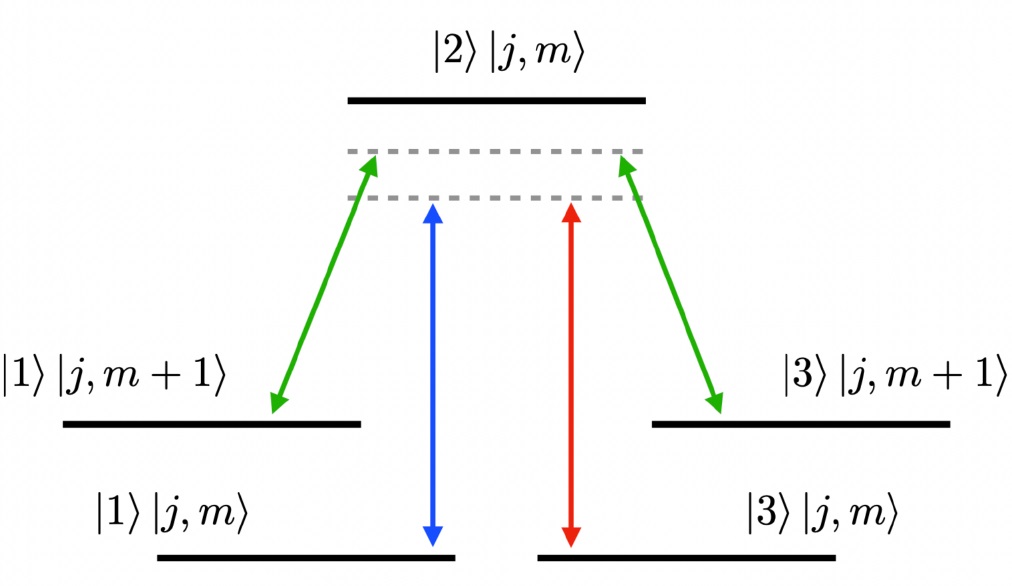}
\end{center}
\caption{$\Lambda$-coupling scheme in the presence of interaction with the spin bath. The state $\Ket{e}\Ket{j,m}$ is coupled to the states $\Ket{g_1}\Ket{j,m}$ and $\Ket{g_2}\Ket{j,m}$ via the two pulses (represented by the blue solid arrow and red solid arrow), and it is coupled also to the states $\Ket{g_1}\Ket{j,m+1}$ and $\Ket{g_2}\Ket{j,m+1}$ through the bath (represented by the green dashed arrows).} \label{fig:Lambda_scheme_and_bath}
\end{figure}

We can now reconsider the coupling scheme, which is better represented in Fig.~\ref{fig:Lambda_scheme_and_bath}, where the state $\Ket{e}\Ket{j,m}$ is shown to be coupled with the states $\Ket{g_k}\Ket{j,m}$ through the pulses, but also with the states $\Ket{g_k}\Ket{j,m+1}$ through the system-environment interaction. A more complete description of the coupling scheme should include also the coupling through the pulses of states $\Ket{g_k}\Ket{j,m+1}$ to $\Ket{e}\Ket{j,m+1}$ and so on.

The simplified notation used for the states of the bath, involving states $\Ket{j, m}$, does not put in evidence the fact that the single-spin angular momenta and the intermediate squared angular momenta are also necessary to identify a quantum state. The generic state $\Ket{j, m}$ is then to be meant as $\Ket{j, m, \{j_1, j_2, ... j_N, j_{12}, j_{123}, ...\}}$, with $j_{1...k} $ the quantum number associated to the square of the angular momentum operator $J_{1...k}^2 = \left(\sum_{i=1}^k \mathbf{J}_i\right)^2$~~\cite{ref:Weissbluth}. Since the squares of all such intermediate angular momenta commute with the total Hamiltonian, they essentially do not play any role in the dynamics, in the sense that, starting from a generic state $\rho_S \otimes  \sum_{mm'} \alpha_{mm'}^{j} \Ket{j, m, \{j_\alpha\}} \Bra{j, m', \{j_\alpha\}}$, the system will evolve into a linear combination of states with the same set of intermediate angular momenta: $\sum_{mm'kk'} c_{jmm'}^{kk'}(t) \Ket{k}\Bra{k'} \otimes \Ket{j, m, \{j_\alpha\}} \Bra{j, m', \{j_\alpha\}}$, where the coefficients $c_{jmm'}^{kk'}(t)$ do not depend on the set of quantum numbers $\{j_\alpha\}$.


\section{Analysis of the model}\label{sec:theoretical}

Two important behaviors of the STIRAP efficiency can be predicted under the assumption of zero-temperature environment, which means assuming the bath initially in the state $\Ket{\{\downarrow\}}$, indicating all the spins in the state $\Ket{\downarrow}$. The first property is the robustness of the population transfer in the weak coupling regime. The second one is the diminishing of efficiency at strong coupling due to the rise of a Zeno partitioning of the Hilbert space.
 
To prove the first property, consider that the generic state $(c_1 \Ket{g_1} + c_2 \Ket{g_2}) \Ket{\{ \downarrow \}}$ belongs to the kernel of $\tilde{H}_E+\tilde{H}_I$ for any values of $c_1$ and $c_2$, which guarantees that it is protected from the environment. Therefore, adiabatic following of the dark state of the STIRAP Hamiltonian is insensitive to the spin bath, as far as diabatic transitions are not significant (such transitions are always present, since in every regime the adiabatic evolution is an approximated one). 

In the strong coupling regime, the efficiency approaches zero. Though the most natural explanation of this fact could seem the dissipation induced by the environment, it is important to note that this is not the case. In fact, the true phenomenon responsible for the diminishing of the efficiency is the formation of Zeno subspaces and the establishment of a Zeno dynamics~\cite{ref:Militello2001,ref:Facchi2002}. Such an effect occurs when a strong interaction is able to render other interactions ineffective. More naively, this happens because the strong interaction can be considered the main Hamiltonian term, whose eigenspaces are the Zeno subspaces, while the other Hamiltonian terms are treatable as perturbations, almost unable to induce transitions between the Zeno subspaces. The relevant constrained dynamics is called Zeno dynamics~\cite{ref:PascaFacchiRev,ref:PascaFacchiMarmoRev}, which can happen to be established even in the presence of non-Hermitian Hamiltonian~\cite{ref:MiliNapo2020}. In our specific case, a very strong system-environment coupling (playing the role of the main Hamiltonian terms) makes the state $\Ket{e}\Ket{\{\downarrow\}}$ to belong to a Zeno subspace different from that to which belong the states $\Ket{g_1}\Ket{\{\downarrow\}}$ and $\Ket{g_2}\Ket{\{\downarrow\}}$, which effectively deactivates the coherent coupling associated with the pulses.
The neutralization of the effects of the pulses can be better understood in terms of the Hamiltonian eigenstates. Consider first that in the strong-coupling regime $\tilde{H}_S$ can be considered as a perturbation with respect to the other terms.
Next, observe that the operator $\tilde{H}_I + \tilde{H}_E$ commutes with the number operator $\hat{N} = \Ket{e}\Bra{e} + \sum_k (\sigma_z^{(k)}+1_k)/2$, which means that a series of invariant subspaces corresponding to specific total excitations can be identified. Concerning the eigenspace with zero excitations, it is spanned by the two ground states $\Ket{g_m}\Ket{\{\downarrow\}}$, with $m=1,2$. The subspace with one excitation has dimension $2L+1$ and is spanned by the following states: $\Ket{e}\Ket{\{\downarrow\}}
$, $\Ket{g_1}\Ket{\{\downarrow\}\uparrow_k\{\downarrow\}}$ and $\Ket{g_2}\Ket{\{\downarrow\}\uparrow_k\{\downarrow\}}$, where $\Ket{\{\downarrow\}\uparrow_k\{\downarrow\}}$ is a state of the bath with all the spin in the state $\Ket{\downarrow}$ except for the $k$-th spin, which is in the state $\Ket{\uparrow}$. 
Let us now introduce the states 
$\Ket{\varphi_m} = \Ket{g_m}\sum_k \xi_k^{(m)} \Ket{\{\downarrow\}\uparrow_k\{\downarrow\}}$, with $\xi_k^{(m)} = \eta_k^{(m)}/\eta_m$ and 
\begin{equation}\label{eq:eta_m}
\eta_m = \sqrt{\sum_k (\eta_k^{(m)})^2} \,.
\end{equation} 
The restriction of the interaction in \eqref{eq:HI-tilde} term to the subspace with one excitation takes the following form:
\begin{eqnarray}
\tilde{H}_I &=&\sum_m \eta_m ( \Ket{e}\Ket{\{\downarrow\}} \Bra{\varphi_m} + \Ket{\varphi_m}\Bra{\{\downarrow\}}\Bra{e}) \,.
\end{eqnarray}
On the other hand, the restriction of $\tilde{H}_E$ given by \eqref{eq:HE-tilde} on the subspace with one excitation has $\Ket{e}\Ket{\{\downarrow\}}$ as eigenstate corresponding to the eigenvalue zero, and all the other states spanning a subspace corresponding to the eigenvalue $\Delta_E$.
Therefore it is straightforward to find that $\tilde{H}_I+\tilde{H}_E$ possesses the following two eigenstates and relevant eigenvalues:
\begin{eqnarray}
\Ket{\Phi_\pm} = \cos\phi(\sin\chi \Ket{\varphi_1} + \cos\chi \Ket{\varphi_2}) \pm \sin\phi \Ket{e}\Ket{\{\downarrow\}} , \quad 
\end{eqnarray}
with
\begin{eqnarray}
\nonumber
\chi &=&\arctan(\eta_1/\eta_2)\,,\\
\nonumber
\phi &=& \arctan\left( \gamma \pm \sqrt{1+\gamma^2}  \right) \,, \\
\nonumber
\gamma &=& \frac{\Delta_E}{\sqrt{(\eta_1)^2+(\eta_2)^2}}  \,,
\end{eqnarray}
and
\begin{eqnarray}
E_\pm = \frac{1}{2} \left(  \Delta_E \pm \sqrt{(\eta_1)^2 + (\eta_2)^2 + \Delta_E^2 }  \right) \,.
\end{eqnarray}
Moreover, there are $2L-1$ eigenstates related to the eigenvalue $\Delta_E$ (each of them has one excitation in the bath), one of which is the state $\Ket{D} = \cos\chi \Ket{\varphi_1} - \sin\chi\Ket{\varphi_2}$. It is crucial to emphasize that $\Ket{e}\Ket{\{\downarrow\}}$ is involved only in the states $\Ket{\Phi_\pm}$.
Once the system Hamiltonian $\tilde{H}_S$ in \eqref{eq:HS-tilde} is added, considering the strong coupling regime, we can treat the effects of this contribution through perturbation theory. Since the  states $\Ket{g_m}\Ket{\{\downarrow\}}$ are connected by $\tilde{H}_S$ only to $\Ket{e}\Ket{\{\downarrow\}}$, the expressions of the relevant eigenstates of $\tilde{H}_S+\tilde{H}_E+\tilde{H}_I$ corrected to the first order are:
\begin{eqnarray}
\Ket{G_m} = \Ket{g_m}\Ket{\{\downarrow\}} + \sum_{\alpha=\pm} \frac{   \Bra{\Phi_\alpha} H_S \Ket{g_m}\Ket{\{\downarrow\}}  }{E_\alpha} \Ket{\Phi_\alpha} \,.
\end{eqnarray}

Under the assumption $\sqrt{(\eta_1)^2 + (\eta_2)^2} \gg |\Delta_E|, |\Omega_0|$ the denominators of the first-order corrections become essentially equal to $\pm\sqrt{(\eta_1)^2 + (\eta_2)^2} / 2$ and the coefficients attributed to $\Ket{\Phi_\pm}$ turn out to be of the order of $\Omega_0/\sqrt{(\eta_1)^2 + (\eta_2)^2}$. Therefore, the higher the coupling constants between the three-state system and the spin bath, the more the states $\Ket{G_m}$ approach $\Ket{g_m}\Ket{\{\downarrow\}}$ at any time, neutralizing the interaction induced by $\tilde{H}_S$. It is anyway important to note that high values of $\Delta_E$ can determine a diminishing of the absolute value of one of the denominators, then compromising the freezing of the dynamics. 
Finally, we remark that for the homogeneous model the condition to satisfy for the Zeno dynamics to occur reduces to $\eta\sqrt{2L} \gg \Delta_E, \Omega_0$.


\section{Numeric Results}\label{sec:numerics}

Here we show some numerical results (obtained solving the Schr\"odinger equation using a fourth-order Runge-Kutta method) for the efficiency of the population transfer in the presence of interaction with the spin environment for the homogeneous model and under the zero-temperature assumption. The latter hypothesis means that it is populated only the ground energy level having all the spins in the state $\Ket{\downarrow}$, which corresponds to $j=L/2$, $m=-j$. 
The three-state system starts in the state $\Ket{g_1}$ and from $t=-T$ to $t=T$ is subjected to the pulses which are supposed to realize the adiabatic following to $\Ket{g_2}$. The relevant quantity defining the efficiency of the process is then nothing but the population of state $\Ket{g_2}$ in the final state $\rho(T)$ of the system:
\begin{equation}
P(T)\equiv\Bra{g_2}\rho_S(T)\Ket{g_2}\,.
\end{equation}
\begin{figure}[h!]
\begin{center}
\begin{tabular}{cc}
\includegraphics[width=0.45\textwidth, angle=0]{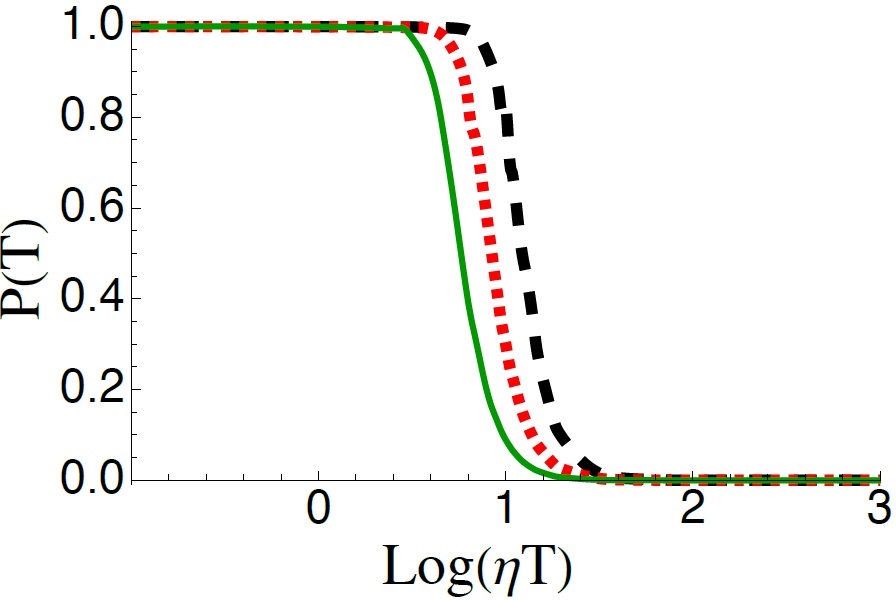} &  
\includegraphics[width=0.45\textwidth, angle=0]{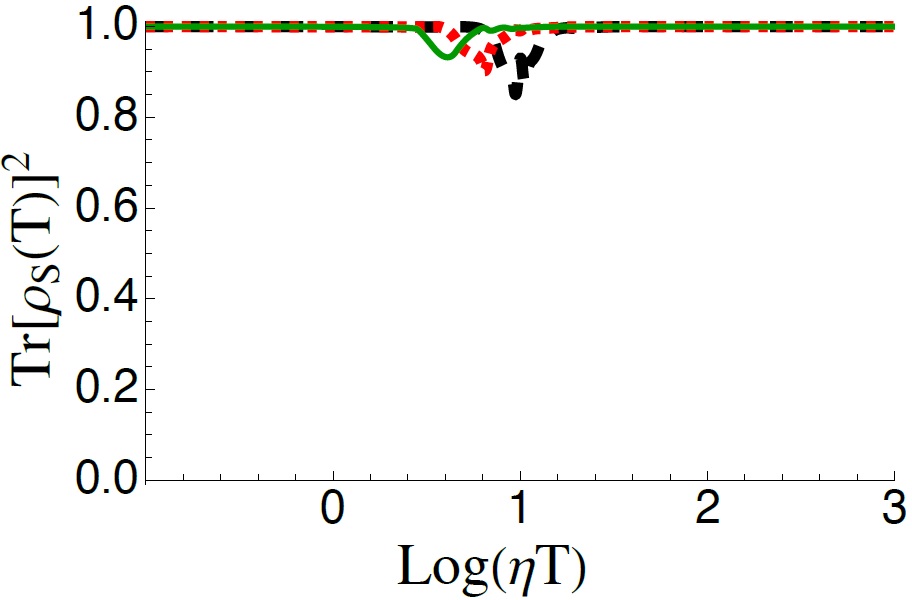} \\ 
(a) & (b)
\end{tabular}
\vskip0.5cm
\begin{tabular}{c}
\includegraphics[width=0.45\textwidth, angle=0]{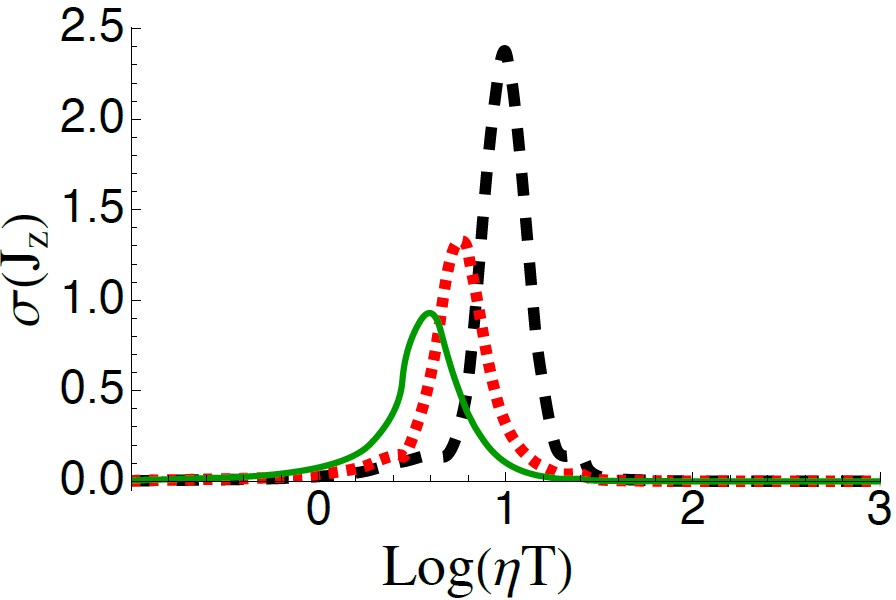} \\
(c)
\end{tabular}
\end{center}
\caption{Dynamical properties for different bath sizes when the system is prepared in state $\Ket{g_1}$. In (a) final population of state $\Ket{g_2}$ ($P(T)\equiv\Bra{g_2}\rho_S(T)\Ket{g_2}$), in (b) purity of the final state of the system and in (c) the variance of $J_z$ of the end of the process ($t=T$), denoted as $\sigma^2(J_z)$. All such quantities are plotted as functions of the coupling constant $\eta$ (in units of $1/T$ and in logarithmic scale). In all the plots  different values of the number of spins in the bath are considered: $L=10$ (black dashed line), $L=20$ (red pointed line), $L=40$ (green solid line). All the other parameters are always the same: $T \Omega_0 = 100$, $\tau/T = 0.1$, $T \Delta_S = 1$, $T \Delta_E = 1$.} 
\label{fig:pop_no_rescaled}
\end{figure}

In Fig.~\ref{fig:pop_no_rescaled}a it is plotted the population of state $\Ket{g_2}$ at time $t=T$ as a function of the coupling constant with the spins of the environment $\eta$. Different sizes of the environment are considered, i.e., different numbers of spins in the environment ($L=10,20,40$). The pulses have specific features which guarantee the validity of the adiabatic approximation (according to the previous discussion), hence allowing for a perfect population transfer, in the ideal case (i.e., in the absence of environment). Indeed, for very small values of $\eta$ the population transfer is perfect, since the final population of $\Ket{g_2}$ is equal to $1$. On the contrary, for higher values of $\eta$ the efficiency diminishes, reaching the value zero.
It is well visible that an increase of the number of spins of the environment determines a diminishing of the efficiency of the population transfer, in the sense that the knee of the curve of efficiency starts for smaller values of $\eta$. In all plots of Fig.~\ref{fig:pop_no_rescaled} the detunings $\Delta_S$ and $\Delta_E$ are assumed to be much smaller than the pulse amplitudes (in particular $1 = T \Delta_S = T \Delta_E \ll T\Omega_0 = 10$), which produce essentially the same results obtained for $\Delta_S=\Delta_E=0$ (whose relevant plots are not reported here).
In Fig.~\ref{fig:pop_no_rescaled}b the relevant purity is reported, showing that the final state is essentially pure except that for the values of $\eta$ corresponding to the slope. In the case of complete population transfer the purity of the state of the three-state system is clearly equal to $1$, since it obviously coincides with $\rho_S(T)=\Ket{g_2}\Bra{g_2}$. Then purity becomes lower in a range of values corresponding to the slope of the efficiency curve and becomes again equal to $1$ when efficiency is zero and the final state is $\rho_S(T)=\Ket{g_1}\Bra{g_1}$.
Fig.~\ref{fig:pop_no_rescaled}c shows the variance of $J_z$ of the environment, in order to establish the involvement of the states of the bath at the end of the population transfer process. 
The higher $\sigma^2(J_z) \equiv \langle J_z^2\rangle - \langle J_z \rangle^2$, the bigger the number of states of the bath involved in the universe state after the action of the pulses. According to one's expectation, this quantity is non vanishing in the same range of values of $\eta$ where the purity is lower than unity.

\begin{figure}[h!]
\begin{center}
\includegraphics[width=0.45\textwidth, angle=0]{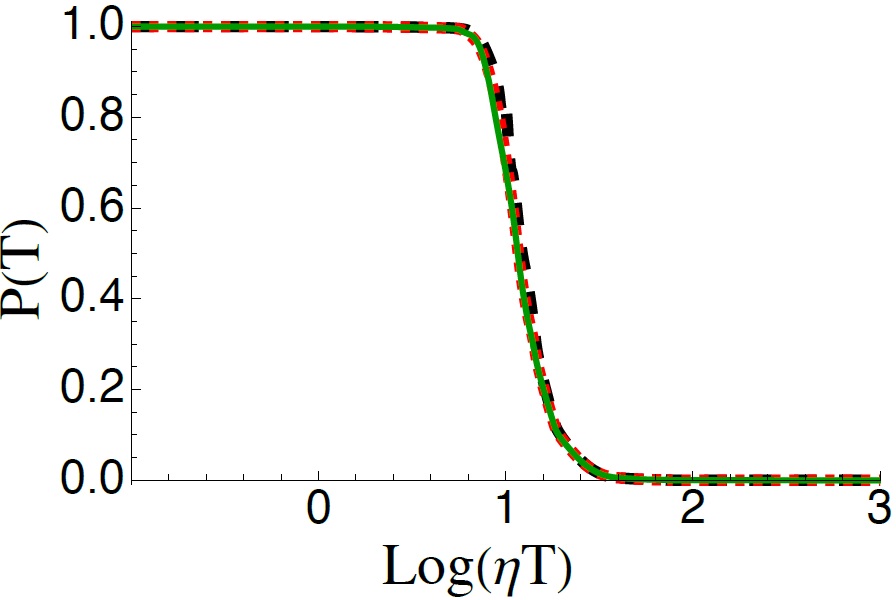}
\end{center}
\caption{Final population of state $\Ket{g_2}$ ($P(T)\equiv\Bra{g_2}\rho_S(T)\Ket{g_2}$) when the system is prepared in the state $\Ket{g_1}$ as functions of the coupling constant $\eta$ (in units of $1/T$ and in logarithmic scale). Different values of the number of spins in the bath are considered and relevant rescaling of the system-environment interaction term by the factor $\sqrt{10/L}$. The spin numbers are: $L=10$ (black dashed line), $L=20$ (red pointed line), $L=40$ (green solid line). All the other parameters are always the same: $T \Omega_0 = 100$, $\tau/T = 0.1$, $T \Delta_S = 1$, $T \Delta_E = 1$. According to the Holstein-Primakoff theory, the three situations are essentially equivalent and the relevant curves consequently coincide.}\label{fig:pop_rescaled}
\end{figure}

According to the general theory of Holstein-Primakoff, there is an effective equivalence between a giant spin (here corresponding to the sum of all the spins of the environment) and a harmonic oscillator~\cite{ref:HolsteinPrimakov}, in the sense that, for a fixed value of the quantum number $j$ associated to $J^2$, restricting to the lowest part of the spectrum of $J_z$, the three operators $J_z + j {1\!\!1}$, $J_+/\sqrt{2j}$ and $J_-/\sqrt{2j}$ act very similarly to the operators $a^\dag a$, $a^\dag$ and $a$ of a one-dimensional harmonic oscillator. All this considered, in Fig.~\ref{fig:pop_rescaled} it is reported the efficiency of the population transfer including the rescaling of the coupling constant (differently from Fig.~\ref{fig:pop_no_rescaled}a), where it is well visible that the three situations considered corresponding to $L=10,20, 40$ produce equivalent results, as a consequence of the fact that they all correspond to the interaction with the same effective harmonic oscillator. For the sake of readability, we remind here that we haven't put the factor $1/\sqrt{L}$ in the model. Therefore, after taking $L=10$ as the reference case, the rescaling in the cases $L=20,40$ is obtained multiplying the system-bath interaction term by the factor $\sqrt{10/L}$, as indicated in the caption.

\begin{figure}[h!]
\begin{center}
\includegraphics[width=0.45\textwidth, angle=0]{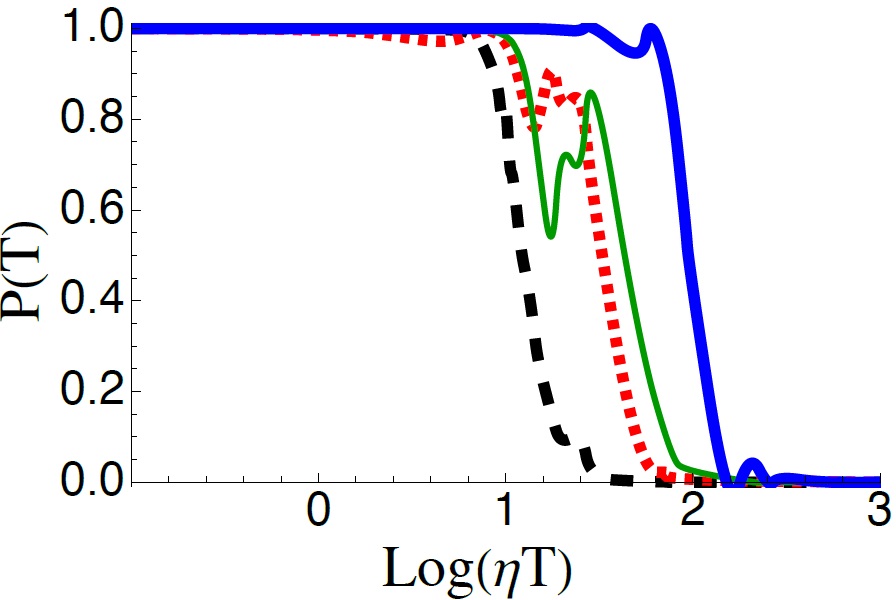}
\end{center}
\caption{Final population of state $\Ket{g_2}$ ($P(T)\equiv\Bra{g_2}\rho_S(T)\Ket{g_2}$) when the system is prepared in the state $\Ket{g_1}$ as functions of the coupling constant $\eta$ (in units of $1/T$ and in logarithmic scale). Different values of $\Delta_E$ are considered: $T \Delta_E = 1$ (black dashed line), $T \Delta_E = 50$ (red pointed line), $T \Delta_E = 100$ (green solid line) and $T \Delta_E = 500$ (blue bold solid line). All the other parameters are always the same: $T \Omega_0 = 100$, $\tau/T = 0.1$, $T \Delta_S = 1$. 
}\label{fig:pop_offresonance}
\end{figure}

In Fig.~\ref{fig:pop_offresonance} it is shown the efficiency of the population transfer for different values of the parameter $\Delta_E$ and for $L=10$ spins. It is important to note that values of $\Delta_E$ very different from $\Delta_S$ imply that the system-environment interaction is off-resonant (in fact, $\Delta_E - \Delta_S = \omega - \nu$), diminishing the effects of the environment on the system. Indeed, for higher values of $\Delta_E$ the knee of the efficiency curve starts for higher values of $\eta$. Anyway, for large enough values of the system-environment coupling constant the efficiency still vanishes.  
It is interesting to note the non-monotonic behavior of the efficiency in the intermediate regime, especially in the two cases $T\Delta_E = 5$ and $T\Delta_E = 10$. The origin of this non-monotonicity is probably traceable back to the competition of effects due to high value of $\eta$ and $\Delta_E$ discussed in the end of sec.~\ref{sec:theoretical}.  

\section{Discussion}\label{sec:discussion}

A model for the STIRAP protocol in the presence of interaction between the three-state system and a spin bath has been introduced and analytically studied, focusing on the case where the bath is initially in its ground state. Behaviors at weak and strong couplings have been predicted.
The numeric resolution of the dynamics (all performed through a fourth-order Runge-Kutta method through a code implemented by us) specialized to the homogeneous case and consequent curves of efficiency of the population transfer vs system-environment coupling constant $\eta$ show in a clear way that the efficiency is essentially equal to $1$ for small values of $\eta$, as one reasonably expects, and zero for high values of such parameter (which is well visible in Fig.~\ref{fig:pop_no_rescaled} and subsequent). The first behavior is in fact intuitive since for very small values of $\eta$ the environment is supposed not to influence significantly the STIRAP process. The second feature seems to be pretty intuitive, too, since in the presence of a strong interaction with the environment the adiabatic following of the eigenstates of the system Hamiltonian is supposed to be jeopardized by dissipation. Nevertheless, it has been shown that the true motivation for such diminishing of the efficiency is to be traced back to the rise of Zeno subspaces and relevant Zeno dynamics, sometimes also referred to as dynamical decoupling (i.e., neutralization of the couplings associated to the pulses) induced by the environment.  In terms of properties of the Hamiltonian, it is possible to show that the states $\Ket{g_k}\Ket{\{\downarrow\}}$ approach two ground states of the total Hamiltonian of system and environment in the regime of very high values of $\eta$, while the state $\Ket{e}\Ket{\{\downarrow\}}$ belongs to a different subspace, which effectively implies neutralization of the pulses supposed to induce the adiabatic population transfer.  In order to further support this interpretation and exclude the occurrence of dissipation in the regime of very strong coupling, consider that if dissipation occurred at high values of $\eta$, since the state $\Ket{e}$ is coupled both to $\Ket{g_1}$ and $\Ket{g_2}$ through the bath, the effect would be a redistribution of the total population between the two ground states, instead of a vanishing population for state $\Ket{g_2}$.
In passing, it is interesting to note that our numeric results confirm the Holstein-Primokoff theory, according to which no significant effects should be observed with an increase of the size fo the environment and simultaneous rescaling of the coupling constant by a factor proportional to the inverse of the number of bath spins. This has been clearly shown in Fig.~\ref{fig:pop_rescaled}. Therefore, in further resolutions,  we have focused on the $L=10$ case in order to reduce the time of calculation.

In the case of spins off-resonant with respect to the transition frequencies of the three-state system, the numeric analysis reported in Fig.~\ref{fig:pop_offresonance} shows again the predictable behaviors for small and very high values of $\eta$, but, in the range of intermediate-high values of $\eta$, there are oscillations of the efficiency. Such behavior is probably due to a complicated interplay between high values of the detuning $\Delta_E$, which tends to deactivate the environment, and the strengthening of the interaction with the spins of the bath.

Concluding, we have predicted the efficiency of a STIRAP protocol of a system interacting with a spin environment, showing that the efficiency is very high for small values of the coupling constant and, because of a Zeno effect, it goes to zero for very high values of such coupling strength. Moreover we have singled out the presence of oscillations of the efficiency in some range of the parameters. The obtained results suggest that it could be interesting to investigate in depth the reason for the oscillatory behavior, which is unexpected because it is seemingly absent in the case of bosonic bath. Since our analysis is based on the conservation of a quantity traceable back to the RWA, it could be interesting to investigate the system without such an approximation. Finally, interactions between the spins of the bath will likely change the predictions here reported.

\section*{Acknowledgments}

The authors gratefully acknowledge financial support from the FFR2021 grant from the University of Palermo, Italy.

\end{document}